# Factoring integers with Young's *N*-slit interferometer

John F. Clauser
*Physics Department, University of California*
*Berkeley, California 94720*

and

Jonathan P. Dowling
*Weapons Sciences Directorate, AMSMI-RD-WS-ST*
*Research, Development, and Engineering Center*
*U. S. Army Missile Command*
*Redstone Arsenal, Alabama 35898-5248*

**ABSTRACT**

We show that a Young's *N* slit interferometer can be used to factor the integer *N*. The device could factor four- or five-digit numbers in a practical fashion. This work shows how number theory may arise in physical problems, and may provide some insight as to how quantum computers can carry out factoring problems by interferometric means.

Recently, Peter Shor has produced a remarkable proof that quantum computers can be used to factor a large integer $N$ using only "polynomial resources" [1]. Such a computer, if built, would have great impact on the fields of public-key encryption and computing in general [2]. Given an $M$-bit, parallel representation of $N$, Shor achieves these results by considering a $M$-particle quantum system that is in a number state, and then by using superposition and interference after unitarily transforming this state into an entangled state. Since all of these ingredients (except entangled states) are available in classical interferometry, we wondered to what extent it might be possible to factor integers by classical interferometric means. Herein, we show that it is indeed possible to factor the integer $N$ using a familiar Young's $N$ slit classical interferometer. In such an interferometer, when $N$ is a product of two integers $n$ and $r$, we show that a simple analysis of the resultant diffraction pattern, and its dependence on illumination wavelength $\lambda$, can be used to determine $n$ and $r$. Our result illustrates an interesting appearance of number theory in physics, and indicates how problems such as factoring might be attacked by novel physical — as opposed to algorithmic — methods. Of course, our classical scheme does not exhibit the exponential decrease in computational resources that apparently can only be obtained on a quantum device. Nevertheless, it is instructive from a fundamental-concept point of view.

Consider a simplification (see Fig. 1, inset) of the 2D problem considered by Clauser and Reinsch (CR) [3]. A planar, finite, $N$ period diffraction grating (in the $xy$ plane at $z = 0$), is illuminated by parallel monochromatic light propagating in the $+ z$ direction. It produces a diffraction pattern on a planar image screen, located on an $xy$ plane at $z = R$. The grating is centered on the $x = 0$ axis and is composed of an odd number $N$ of parallel slits with width $s$ and period $a$. (If $N$ is even, then 2 is a factor.) We take the transmission function $s(x)$ for a single slit of the grating (defined on the interval, $a/2 \leq x \leq a/2$) to be 1 for $s/2 \leq x \leq s/2$, and zero otherwise. Furthermore, we choose the illumination wavelengths $\lambda_n$ such that $\lambda_n = a^2 n/R$, where $n$ is an integer. Since we are searching for factors of $N$, and $N$ is odd, then for $n$ to be a factor, $n$ must also be odd. For the above conditions, *and* when $n$ is a factor of $N$, CR have shown that the wave amplitude on the screen is given by [3]

$$\psi_{n,N}(x) \propto \int_{-\infty}^{+\infty} E_n(v) P_r\left(v - \frac{1}{2}\right) t'_n(x-av) dv, \tag{1}$$

where

$$t'_n(x) \equiv \frac{1}{n} \sum_{q=(1-n)/2}^{(n-1)/2} s(x - a\,q) \tag{2}$$

represents the transmission of a "shortened" modified $n$-period grating. We have defined the functions $E_n(v) \equiv \exp(i\pi v^2/n)$ and $P_r(v) \equiv \sin(\pi r v)/\sin(\pi v)$, where $r \equiv N/n$ is also an odd integer, since $N$ is odd and $n$ is one of its factors.

The above formulae give an exact result, at least within the limits of validity of the Fresnel approximation and the Kirchoff diffraction integral. The function $P_r(v)$ is periodic with period one. It consists of a string of positive spikes of height $r$ centered on integer values of $v$. In the limit as $N$ and $r$ approach infinity, then $P_r(v)$ approaches a comb function (infinite periodic string of delta functions), and Eqs. (1) and (2) imply that the diffraction pattern will consist of a "filtered" self-image of the grating, an example of the Talbot effect [3].

A necessary condition for obtaining the above result is that $n$ be a factor of $N$, so that the individual slits, taken $n$ at a time, [via Eq.(2)] add coherently. When additional slits are appended to either end of the grating, such that $n$ is no longer a factor of $N$, then the unique filtered-self-imaging properties of Eq. (1) no longer hold exactly, but instead only approximately. The resulting error then represents a remainder to the factorization.

To illustrate these ideas, let us first consider the idealized limiting case with very high intensity illumination and infinitesimal slit widths. In this limit we can approximate the transmission function (ignoring the normalization) as $s(x) \approx \delta(x)$, whereupon the integral in Eq. (1) may be evaluated as

$$\psi_{n,N}(x) \propto P_r\left(\frac{x}{a} - \frac{1}{2}\right) \sum_{q=(1-n)/2}^{(n-1)/2} \exp\left[\frac{i\pi}{n}\left(\frac{x}{a} - q\right)^2\right]. \qquad (3)$$

Here we have used the fact that the function $P_r$ is periodic, with period one, to move it outside of the summation. The function $P_r$ is peaked at integer values of its argument, i.e., at values of $x/a = l + 1/2$, where $l$ is an integer. Defining the intensity at any position $x$ on the screen as $I_{nN}(x) \equiv \psi_{nN}^*(x)\,\psi_{nN}(x)$, then the intensity at the position of the $l$'th peak is given by

$$\hat{I}(l,n,N) \equiv \psi_{nN}^*\left(l\,a + \frac{a}{2}\right)\psi_{nN}\left(l\,a + \frac{a}{2}\right). \qquad (4)$$

Upon substitution of the above formulae into this equation, we find $\hat{I}(l,n,N) \propto$

$r^2\,\Sigma(l,n)$, where we have defined the functions

$$\Sigma(l,n) \equiv \sum_{p=(1-n)/2}^{(n-1)/2} \sum_{q=(1-n)/2}^{(n-1)/2} f(l,n,p,q), \tag{5}$$

and

$$f(l,n,p,q) \equiv \exp\left\{\frac{i\pi}{n}\left[\left(l+\frac{1}{2}-q\right)^2 - \left(l+\frac{1}{2}-p\right)^2\right]\right\}. \tag{6}$$

Using Eq. (6), and the fact that $p$, $q$, $l$, and $n$ are all integers with $n$ odd, after some algebra, we find the recursion relation, $f(l,n,p+n,q) = f(l,n,p,q)$. Using this relation, it is possible to express the double summation of Eq. (5) as

$$\Sigma(l,n) = \sum_{b=0}^{(n-1)} \sum_{q=(1-n)/2}^{(n-1)/2} f(l,n,b+q,q). \tag{7}$$

Further, using Eq. (6), we have

$$f(l,n,p,q) = \exp\left[\frac{i\pi}{n}(q-p)(q+p-2l-1)\right]. \tag{8}$$

Combining Eqs. (7) and (8), we find

$$\Sigma(l,n) = \left\{ n + \sum_{b=1}^{(n-1)} \exp\left[-\frac{i\pi}{n}b(b-2l-1)\right] \sum_{q=(1-n)/2}^{(n-1)/2} \exp\left(-2i\pi\frac{b}{n}q\right) \right\}. \tag{9}$$

The summation over $q$ in Eq. (9) represents a geometric series which may be summed to yield

$$\Sigma(l,n) = \left\{ n + \sum_{b=1}^{(n-1)} \exp\left[-\frac{i\pi}{n}b(b-2l-1)\right] \frac{\sin(2\pi b)}{\sin\left(2\pi\frac{b}{n}\right)} \right\}. \tag{10}$$

In the ratio of sine functions in Eq. (10), the numerator is always zero, since $b$ is an integer. The ratio correspondingly will vanish for all $b$ unless the denominator likewise vanishes for some values of *b*. However, the summation over *b* always has $b/n < 1$ and $b/n \neq 1/2$ (since *n* is odd), so that the denominator never vanishes for any value of *b*. Thus, the whole sum over *b* in Eq. (10) vanishes; $\Sigma(l,n) = n$ is then just a constant. The resulting intensity is then given by $\hat{I}(l,n,N) \propto r^2 n = N^2/n$ for all values of *l*. That is, all spikes in the diffraction pattern have the same height! Moreover, this result does not obtain when the wavelength is chosen so that *n* is an integer, but *not* one that is a factor of *N*. Thus, one can determine if *n* is not a factor of *N* by simply looking at the diffraction pattern produced by illumination at various wavelengths $\lambda_n$, chosen so that $\lambda_n \equiv a^2 n/R$ produces an odd integer value for *n*.

To illustrate these results, we calculate numerically the diffraction pattern produced under the above conditions with *n* odd but not necessarily a factor of *N*. The intensity is then given by

$$I_K(x,n,N) \propto \left| \int_{-\infty}^{\infty} \sum_{q=(1-N)/2}^{(N-1)/2} s(\xi - a q) \exp\left[ \frac{i\pi}{\lambda_n R} (\xi-x)^2 \right] d\xi \right|^2 . \quad (11)$$

We consider five different cases wherein the number of slits is a product of two primes, i.e., gratings with $N = 55 = 5 \times 11$, $N = 95 = 5 \times 19$, $N = 119 = 7 \times 17$, $N = 141 = 3 \times 47$ and $N = 143 = 11 \times 13$ slits, respectively, one case wherein the number of slits is prime, i.e., $N = 139$, and one case wherein the number of slits is a product of three primes, i.e. $N = 105 = 3 \times 5 \times 7$.

The parameters used in the simulations are $R = 10$ m, $a = 1$ cm. The wavelength is varied in steps [4], consistent with $n$ being an odd integer, for values of $n = 1,N$, as per $\lambda_n = a^2 n/R$.

Figures 1a, 1b, and 1c show the intensity distribution as a function of $x$ for $N = 143$, and $n = 11, 13,$ and $17$, respectively. Indeed, the cases wherein $n$ is a factor of $N$, with $n = 11$ and $13$, show all spikes with the same height, while the pattern with $n = 17$, wherein $n$ is not a factor of $N$, shows spikes with differing heights.

At the $l$ 'th spike's peak, the intensity [via Eq. (11)] will be given by $\hat{I}_K(l,n,N) \equiv I_K(la + a/2, n, N)$. To determine the magnitude of the variation of spike height as a function of $n$, we evaluate the RMS variation of $\hat{I}_K(l,n,N)$ over spikes $l = 1$ to $(N–1)/2$, where we define this variation $\hat{\sigma}$ by the simple formulae

$$\bar{I}(n,N) \equiv \frac{2}{N-1} \sum_{l=1}^{(N-1)/2} \hat{I}_K(l,n,N), \tag{12}$$

and

$$\hat{\sigma}(n,N) \equiv \left[ \frac{2}{N-1} \sum_{l=1}^{(N-1)/2} \left(1 - \frac{\hat{I}_K(l,n,N)}{\bar{I}(n,N)}\right)^2 \right]^{1/2}. \tag{13}$$

Figures 2a – 2e show the dependence of $\hat{\sigma}(n,N)$ upon $n$ for the cases, $N = 55, 95, 119, 141,$ and 143, respectively. For comparison, Figure 2f shows this dependence for the case with a prime number of slits, $N = 139$, while Figure 2g shows this dependence for the case $N = 105$. As anticipated above, we note that the RMS variation vanishes when $n$ is a divisor (factor) of $N$, but does not when it is not.

In contrast to the above results, when finite width slits are used, the amplitude produced by a single slit then diminishes for large diffraction angle. As such it will not illuminate the whole diffraction pattern of width $Na$ [3], but instead only the portion within its own diffraction pattern, whose half-width on the image plane will be $(\Delta x)_{1/2} \approx \lambda R/(2s)$. One is then led to expect that [3] the ability of the $N$ slit interferometer to factor $N$ will fail unless $sN/(an) \lesssim 1/2$, or equivalently, $s/a \lesssim n/(2N) = 1/(2r)$ holds.

To test this proposition, we define the average intensity of the $l$'th single-slit self-image by

$$I_s(l,n,N) \equiv \frac{1}{s}\int_{-s/2}^{s/2} I_K\left(l a + \frac{a}{2} + \xi, n, N\right)d\xi, \qquad (14)$$

and, exactly as per Eqs. (12) and (13), define an associated mean and RMS variation as $\bar{I}_s(n,N)$ and $\sigma_s(n,N)$, respectively.

We now consider ten of the cases discussed above. Figure 3 shows $\sigma_s(n,N)$ as a function of $s/a$ for these various cases and for $0 \leq s/a \leq 0.15$. Consistent with the main point of this paper, we note that in all cases $\sigma_s(n,N)$ vanishes in the limit as $s$ goes to zero. A further examination of the results of Fig. 3 indicates that with increasing $s/a$, the utility of the diffraction pattern for finding factors of $N$ deteriorates first and most rapidly for the smaller of the two factors of $N$. This deterioration [increase of $\sigma_s(n,N)$] is consistent with the constraint set by $s/a \lesssim n/(2N)$. If the various curves of Fig. 3 are replotted as a function of ($Ns$(/($na$) then they all (approximately) collapse to form a single curve, with a shape similar to that of $n, N$ = 3, 141 of Fig. 3.

To summarize: An important (unsolved) problem in number theory is the question of how to rapidly factor a large integer $N$ into its prime integer divisors [2]. Quantum computation [1] has been offered as a possible solution; however, as yet its practical implementation is wanting. As a first step toward understanding how a novel application of a physical process might be used to factor integers in a nontraditional fashion, we have presented here a classical, optical factoring "engine" that utilizes the ingredients of coherent superposition and interference in order to factor small integers $N$. (With current grating technology, four- to five-digit integers could be easily factored.) Not surprisingly, in our classical device the resources required, such as grating size and illumination power, grow exponentially with the number of bits $M$ needed to represent $N$ as a binary number. This is in contrast to the proposed quantum device, in which resources would scale polynomially with $M$. We conjecture that the key difference is manifest in the quantum entanglement that can be achieved only with a quantum device.

The condition $\lambda_n = a^2 n/R$ specifies that $\lambda$ be tuned via discrete steps. It is noteworthy that the full intensity period-$a$ self-imaging spike patterns of Fig. 1 form at half-period-shifted locations *only* at tuning "resonances" with $\lambda \approx \lambda_n$ with $n$ odd. The allowed tuning errors (and resonance widths) are discussed at length in Ref. 3. At intermediate wavelengths, very complicated interference patterns form. Since our algorithm specifies monitoring the intensity only at narrow neighborhoods centered on the spikes, then an on-resonance condition is readily detected by marginally readjusting the tuning, as necessary, to maximize the average intensity at all spike positions, and/or (with additional detectors) to minimize it elsewhere. So doing, a wavelength-to-grating-period relative calibration is also accomplished.

**ACKNOWLEDGEMENTS**

The authors acknowledge M. Reinsch for contributions, and J. Franson, Shifang Li, and S. Pethel for helpful suggestions. This work was supported by the Office of Naval Research and by the firm, J. F. Clauser & Associates.

**Figure Captions**

Fig. 1. A diffraction grating consisting of $N$ slits of width $s$ and period $a$ is located at $z = 0$, and is illuminated by parallel monochromatic light propagating in the $+z$-direction (inset). It produces a diffraction pattern on a screen, located at $z = R$. Intensity distribution as a function of $x$ for a portion of the diffraction pattern produced by a grating with $N = 143$ infinitesimally-wide slits: (a) $n = 11$, (b) $n = 13$, and (c) $n = 17$.

Fig. 2. Dependence of the RMS variation $\hat{\sigma}(n,N)$ upon $n$ for the cases: (a) N = 55, (b) N = 95, (c) N = 119, (d) N = 141, (e) N = 143, (f) N = 39 (a prime number), and (g) N = 105 (a product of three primes). Zeros indicate factors of $N$, while non-zero dips occur at "near factors".

Fig. 3. Dependence of the RMS variation, $\sigma_s(n,N)$, on s/a. Curves are labeled by $\{n,N\}$.

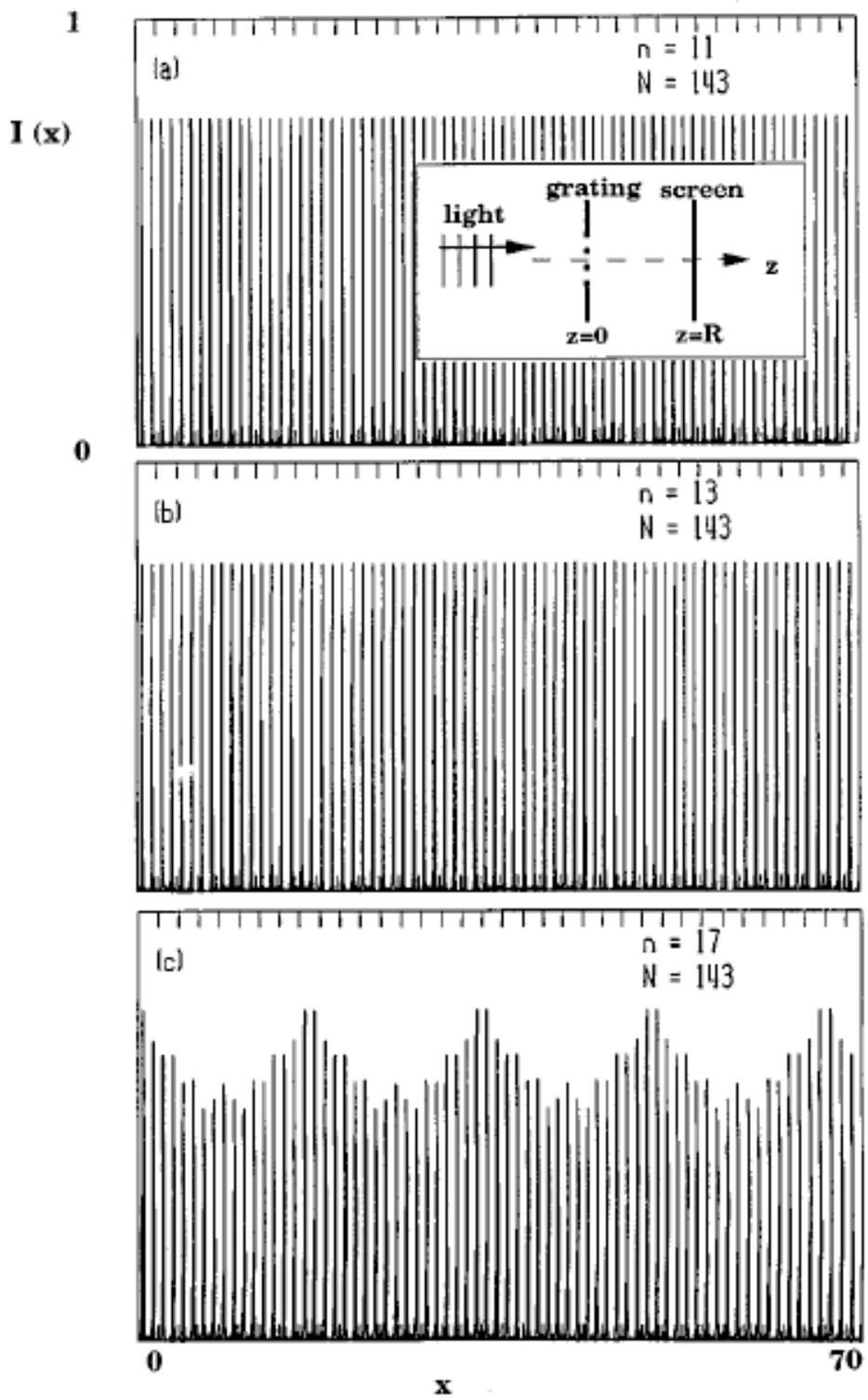

Figure 1

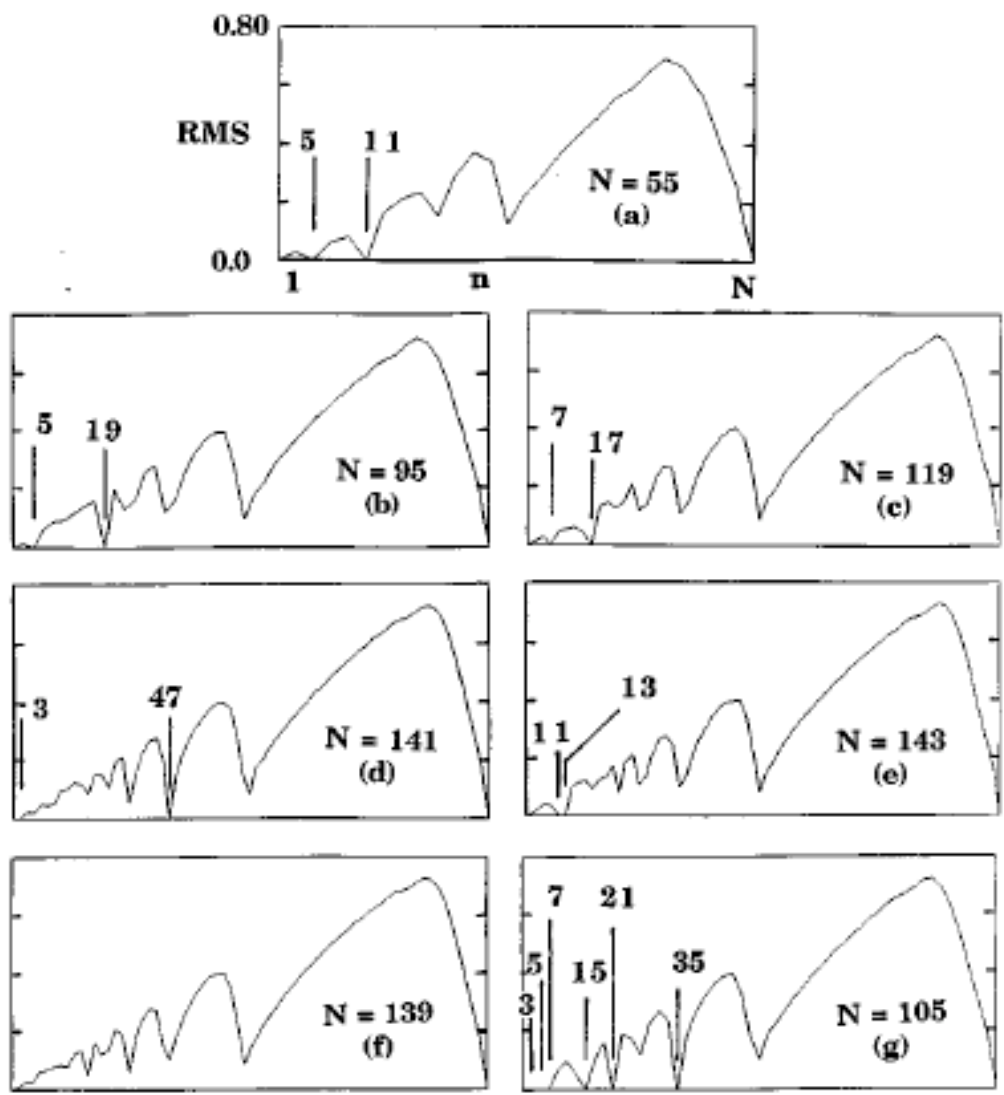

Figure 2

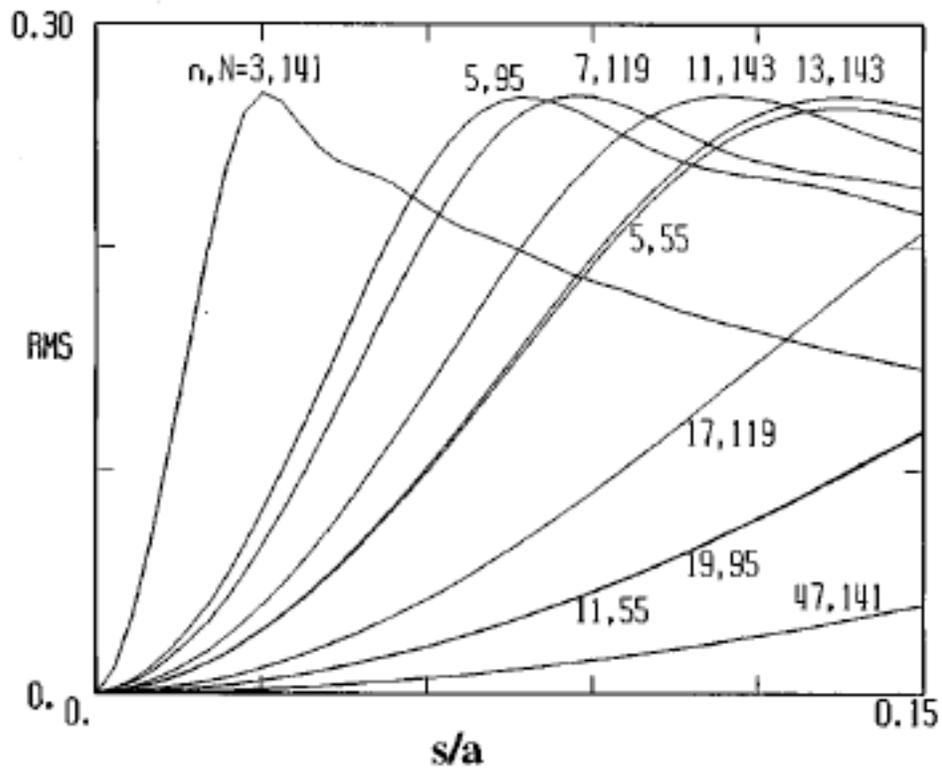

Fig. 3.